\documentclass[12pt, preprint]{aastex}
\begin{document}
\title{THE EXTENDED ZEL'DOVICH MASS FUNCTIONS OF CLUSTERS AND ISOLATED CLUSTERS 
IN THE PRESENCE OF PRIMORDIAL NON-GAUSSIANITY}
\author{Seunghwan Lim\altaffilmark{1}, Jounghun Lee\altaffilmark{2}}
\altaffiltext{1}{Department of Astronomy, University of Massachusetts, LGRT-B 619E, 710 North Pleasant Street, 
Amherst, MA 01003-9305, USA; slim@astro.umass.edu}
\altaffiltext{2}{Astronomy Program, Department of Physics and Astronomy, FPRD, 
Seoul National University, Seoul 151-747, Korea; jounghun@astro.snu.ac.kr}
\begin{abstract}
We present new formulae for the mass functions of the clusters and the isolated clusters with non Gaussian initial 
conditions. For this study, we adopt the Extended Zel'dovich (EZL) model as a basic framework, focusing on the case 
of primordial non-Gaussianity of the local type whose degree is quantified by a single parameter $f_{nl}$.   
By making a quantitative comparison with the N-body results, we first demonstrate that the EZL 
formula with the constant values of three fitting parameters still works remarkably well for the local $f_{nl}$ case. 
We also modify the EZL formula to find an analytic expression for the mass function of isolated clusters which turns 
out to have only one fitting parameter other than the overall normalization factor and showed that the modified EZL 
formula with a constant value of the fitting parameter matches  excellently the N-body results with various values of 
$f_{nl}$ at various redshifts.  Given the simplicity of the generalized EZL formulae and their good agreements with 
the numerical results, we finally conclude that the EZL mass functions of the massive clusters and isolated clusters 
should be useful as an analytic guideline to constrain the scale dependence of the primordial non-Gaussianity of the 
local type.
\end{abstract}
\section{INTRODUCTION}
\label{sec:intro}

The three pillars which have founded and sustained the concordance cosmology are the Cosmic Microwave 
Background (CMB) spectrum, the luminosity-distance relation of the type Ia Supernovae (SNIa), and the statistics of the 
large-scale structures.  The sensational concord among the three observables depicts a simple universe whose initial 
conditions are exhaustively specified by the following six key cosmological parameters: the matter density parameter 
($\Omega_{m}$), the cosmological constant $\Lambda$ parameter ($\Omega_{\Lambda}$), the baryon density 
parameter ($\Omega_{b}$), the dimensionless Hubble parameter ($h$), the amplitude of the linear density power 
spectrum ($\sigma_{8}$), and the spectral index ($n_{s}$) \citep[for a recent review, see][]{hamilton13}. 

The recently reported tensions among the three on the best-fit values of the key parameters, however, have implied that 
the concordance cosmology might be a misnomer. For instance, the Planck CMB analysis yielded the value of $h$ 
to be $0.673\pm 0.012$ \citep{planck_cp13}, which differs substantially from the value, $h=0.738\pm0.024$, 
determined by the HST (Hubble Space Telescope) observation of the SNIa \citep{HST11}. 
A more significant tension with the CMB result was found in the locally determined best-fit value of $\sigma_{8}$: 
The redshift evolution of the SZ (Sunyaev-Zel'dovich) cluster counts traced by the Planck team found 
$\sigma_{8}=0.77\pm0.02$ \citep{planck_sz13} while the best-fit value from the Planck CMB data was 
$\sigma_{8}=0.834\pm0.027$ \citep{planck_cp13}. 

Although our confidence in the concordance cosmology has yet to be shattered down, these discrepancies definitely 
required us  not only to search for possible systematics in observational data analysis but also to carefully reexamine 
whether or not our theoretical modeling of the three observables is accurate enough to predict the initial conditions of
the universe.  Especially, the statistical properties of the large scale structures are hard to model accurately by using
only the first principles since the formation and evolution of the large scale structures occurred in the complicated 
nonlinear regime \citep[e.g., see][]{springel-etal06}. 

The cluster mass function is defined as the number density of the galaxy clusters per mass bin per unit volume. It 
exhibits exponential dependence on the cluster mass, corresponding to the high-mass section of the halo 
mass function. It is one of those few statistics of the large-scale structures which have a direct connection with the initial 
conditions of the universe.  
When \citet{PS74} proposed for the first time an analytic prescription of deriving the halo mass function from the 
initial Gaussian density field, their work could not attract much attention mainly due to its statistical flaw. However, 
ever since \citet{bond-etal91} refined and improved the Press-Schechter formalism with the help of the  celebrated 
excursion set theory, the halo mass function has been the subject of the extensive analytic studies 
\citep[e.g.,][]{jedam95,monaco95,BM96a,BM96b,yano-etal96,
audit-etal97,monaco97a,monaco97b,LS98,SMT01,CL01,ST02,shen-etal06,
des08,MR10a,MR10b,CA11a,CA11b,MS12,paranjape-etal12,PS12,AC12,paranjape-etal13, 
achitouv-etal13a,achitouv-etal13b}.

While the above analytic studies have indeed enlightened us about how to relate the number densities of dark halos to 
the statistical properties of the linear density field and how to account for the effect of environments on the halo mass 
function, it has been gradually realized that the level of the accuracy required for the cluster mass function to be useful 
as a probe of precision cosmology can be achieved only by adjusting the analytic mass functions to the numerical 
results from N-body simulations. The advent of high-resolution N-body simulations and the necessity of having a 
practical but accurate mass function led many authors to come up with empirical formulae, most of which have 
pulled it off to match the numerical results  within $20\%$ errors at the expense of giving up physical understanding 
\citep[e.g.,][]{ST99,jenkins-etal01,warren-etal06,tinker-etal08,crocce-etal10,pph10,courtin-etal11}.  

The extended Zel'dovich (EZL) formula recently proposed by \citet{LL13} is  one of those empirical formulae which 
showed remarkable agreements with various N-body results when their fitting parameters are numerically adjusted.  
Unlike the other formulae, however, the EZL mass function has a good advantage that the empirically 
determined best-fit values of their three characteristic parameters are constant at various redshifts even when the key 
cosmological parameters of the standard $\Lambda$CDM model change.   Given this advantage, \citet{LL13} 
speculated that the EZL mass function of galaxy clusters would provide a tight constraint on the dark energy equation 
of state provided that it is valid even for non-standard cosmologies.  Very recently, \citet{LL14} showed that a modified 
version of the EZL formula is very useful to analytically evaluate the mass function of the superclusters.

As a first step toward testing the validity of the EZL model for non-standard cosmologies, we will investigate here 
whether or not it works even in the presence of primordial non-Gaussianity.  As it has been well known,  primordial 
non-Gaussianity is one of the hottest issues in the field of inflation since detection of  a significant signal of primordial 
non-Gaussianity would rule out the single field slow-roll inflation model 
\citep[for a review, see][]{png_review04}.
Although the degree of primordial non-Gaussianitiy on the CMB scale has been found to be almost undetectably 
small \citep{planck_png13}, the Planck results have yet to demolish the mission to search for a signal of primordial 
non-Gaussianity on the cluster-mass scale given that primordial non-Gaussianity could be scale-dependent. 

Plenty of literatures have already studied the effect of primordial non-Gaussianity on the cluster abundance and its 
evolution \citep[e.g.,][]{LM88,matarrese-etal00,verde-etal01,benson-etal02,sco-etal04,loverde-etal08,dalal-etal08,LS09b,
lam-etal09,MR10c,pph10,desimone-etal11,AC12}.  
The most optimal formula for the cluster mass function as a probe of primordial non-Gaussianity, however, should be 
the one which exhibit not only excellent agreements with the numerical results but also insensitivity of its fitting 
parameters to the redshifts, background cosmology and the presence of primordial non-Gaussianity. 
In this Paper, we will also modify the EZL formula to find an accurate formula for the mass function of the isolated 
clusters which is expected to be more sensitive to the presence of primordial non-Gaussianity \citep{SL09,lee12,AC12}
and examine its limitation as well as usefulness as a probe of primordial non-Gaussianity. 

\section{THE EZL MASS FUNCTION OF ALL CLUSTERS}

\subsection{The Original Formula : A Brief Review}\label{sec:ezl}

The halo mass function, $dN(M, z)/d\ln M$, represents the differential number density of the bound halos in a 
logarithmic mass interval of $[\ln M,\ \ln M +d\ln M]$ at redshift $z$ per unit volume. It depends sensitively on the 
background cosmology via its dependence on the rms fluctuation of the linear density field on the mass scale $M$ at 
redshift $z$, $\sigma(M,z)\equiv D(z)\sigma(M,0)$ where $D(z)$ is the linear growth factor that has a value of unity 
at the present epoch \citep[for a recent review, see][]{zentner07}. 

The extended Zel'dovich (EZL) model is an empirical formula for the halo mass function characterized by 
three fitting parameters, recently developed by \citet{LL13}, under the usual assumption that the primordial density 
field is Gaussian random:
\begin{equation}
\label{eqn:ezl}
\int_{C}\Pi_{i=1}^{3}d\lambda_{i}\,p[{\vec\lambda};\sigma(M,z)] = 
\int_{\ln M}^{\infty}d\ln M^{\prime}\frac{M^{\prime}}{{\bar\rho}}\frac{dN(M^{\prime},z)}{d\ln M^{\prime}}
P({\vec\lambda}\ge {\vec\lambda}_{c}|{\vec\lambda}^{\prime}={\vec \lambda}_{c})\, , 
\end{equation}
where three eigenvalues (in a decreasing order) of the linear deformation tensor on the mass scale of $M$ and $M^{\prime}$ are 
denoted as ${\vec \lambda}\equiv (\lambda_{1}, \lambda_{2}, \lambda_{3})$ and ${\vec \lambda}^{\prime}\equiv 
(\lambda^{\prime}_{1}, \lambda^{\prime}_{2},\lambda^{\prime}_{3})$, respectively. 

As mentioned in \citet{LL13}, the EZL model was established in the framework of the Jedamzik formalism \citep{jedam95}. 
In the right hand side of Equation (\ref{eqn:ezl}), $(M^{\prime}/\rho)dN/d\ln M^{\prime}$ (with mean mass density $\rho$) represents 
the differential fraction of the volume of the linear density field occupied by the proto-halo regions where three eigenvalues reach the 
thresholds as ${\vec \lambda}_{c}\equiv (\lambda_{1c}, \lambda_{2c}, \lambda_{3c})$ when the linear density field is smoothed on the 
mass scale of $M^{\prime}$, and  $P({\vec\lambda}\ge {\vec\lambda}_{c}|{\vec\lambda}^{\prime}={\vec \lambda}_{c})$ represents 
the conditional probability of finding a region embedded in the proto-halos regions where the three eigenvalues exceed the thresholds
on some lower mass scale $M\le M^{\prime}$. 

Integration of the differential volume fraction multiplied by the conditional mass function over $\ln M^{\prime}$ in the right-hand side 
of Equation (\ref{eqn:ezl}) excludes the contributions from the clouds-in-clouds to the cumulative probability that the linear shear 
eigenvalues ${\vec \lambda}$ exceed the given thresholds ${\vec \lambda}_{c}$ on the mass scale of $M$ expressed in the left-hand 
side of Equation (\ref{eqn:ezl}):
\begin{equation}
\label{eqn:plambda}
\int_{C}\Pi_{i=1}^{3}d\lambda_{i}\,p[{\vec\lambda};\sigma(M)]= 
\int_{\lambda_{1c}}^{\infty}d\lambda_{1}\,
\int_{\lambda_{2c}}^{\lambda_1}d\lambda_{2}\,\int_{\lambda_{3c}}^{\lambda_{2}}
d\lambda_{3}\,p[{\vec\lambda};\sigma(M)]\, , 
\end{equation}
where $p(\lambda_{1},\lambda_{2},\lambda_{3})$ is the joint probability density distribution of the linear shear 
eigenvalues, which was first derived in the seminal paper of  \citet{dor70}. 
The conditional probability, $P({\vec\lambda}\ge {\vec\lambda}_{c}|{\vec\lambda}^{\prime}={\vec \lambda}_{c})$, in the 
right-hand side of Equation (\ref{eqn:ezl}) can be calculated as 
\begin{equation}
\label{eqn:pmm}
\frac{\int_{\lambda_{1c}}^{\infty}d\lambda_{1}
\int_{\lambda_{2c}}^{\lambda_{1}}d\lambda_{2}
\int_{\lambda_{3c}}^{\lambda_{2}}d\lambda_{3}\
p(\lambda_{1},\lambda_{2},\lambda_{3},
\lambda^{\prime}_{1}=\lambda_{1c},\lambda^{\prime}_{2}=\lambda_{2c},\lambda^{\prime}_{3}=\lambda_{3c})}
{p(\lambda^{\prime}_{1}=\lambda_{1c},\lambda^{\prime}_{2}=\lambda_{2c},\lambda^{\prime}_{3}=\lambda_{3c})}\, .
\end{equation}
The analytic expression for the joint probability density distribution, $p({\vec\lambda}, {\vec\lambda}^{\prime})$, in 
Equation (\ref{eqn:pmm}) has been found in \citet{des08} and \citet{DS08}.  

Similar to the original formula of \citet{jedam95}, the EZL formula is an integro-differential equation which yields an 
automatically normalized mass function. Unlike the original Jedamzik formula, however, it is not a physical 
model but a mere  empirical formula whose three characteristic parameters, ${\vec \lambda}_{c}$,  have to be 
determined empirically via fitting and thus the best-fit values of ${\vec \lambda}_{c}$ have nothing to do with 
a real physical condition for the gravitational collapse. Nevertheless, the best-fit values of the EZL parameters 
were found by \citet{LL13} to be constant against the variation of redshifts and the initial conditions.  

\subsection{Incorporation of the non-Gaussian Initial Conditions}\label{sec:png_ezl}

In the current work we focus on the case of primordial non-Gaussianity of the local type  
where the deviation of the primordial velocity potential field ($\Phi$) from an Gaussian random field ($\phi$)  
is approximated at first oder as $\Phi \approx \phi +f_{nl}(\phi^{2}-\langle\phi^{2}\rangle)$ where $f_{nl}$ 
is called the (local) primordial non-Gaussianity parameter \citep[see,][]{ngshape04,loverde-etal08}.
\citet{lam-etal09} showed that in the presence of primordial non-Gaussianity of the local type the 
probability density distribution of the shear eigenvalues, $p_{\rm ng}({\vec \lambda};\sigma)$, is approximated 
at first order as  
\begin{equation}
\label{eqn:plambda_png}
p_{\rm ng}({\vec\lambda};\sigma)\approx \Big[1+\frac{\sigma S_{3}}{6}H_3\bigg(\frac{\delta}{\sigma}\bigg)\Big]
p({\vec\lambda};\sigma)\, ,
\end{equation}
where the $H_{3}$ is the third order Hermite polynomial given as $H_3(x)=x(x^{2}-3)$ and $p({\vec \lambda};\sigma)$ 
and  is the probability density distribution of the shear eigenvalues for the Gaussian case ($f_{nl}=0$). 

The skewness parameter, $S_{3}$, in Equation (\ref{eqn:plambda_png}), is related to $f_{nl}$, as \citep{loverde-etal08,lam-etal09}
\begin{equation}
\label{eqn:S3}
\sigma S_{3}\equiv \frac{\langle\delta^3\rangle}{\langle\delta^2\rangle^{3/2}}=\frac{2f_{nl}\gamma^3}{\sigma^3}\, .
\end{equation}
Here $\sigma$ and $\gamma$ are given as 
\begin{eqnarray}
\label{eqn:sig_png}
\sigma^2&=&\frac{1}{(2\pi)^3}\int{\frac{dk}{k}4\pi k^7M^2(k)P_{\Phi}(k)W^2(kR)},\\
\label{eqn:gam_png}
\gamma^3&=&\frac{2}{(2\pi)^4}\int{\frac{dk_1}{k_1}k_1^5M(k_1)W(k_1R)}
           \int{\frac{dk_2}{k_2}k_2^5M(k_2)W(k_2R)} \nonumber \\
&&~~~~\times\int{d\mu_{12}k_{12}^2M(k_{12})W(k_{12}R)\frac{B_\Phi(k_1,k_2,k_{12})}{2f_{nl}}} 
\end{eqnarray}
where $M(k)\equiv[3D(z)c^2T(k)]/(5\Omega_{m}H_{0}^2)$ and $T(k)$ is the transfer function.  The power spectrum,   
$P_{\Phi}$,  and the bispectrum, $B_{\Phi}$, are approximated at first order as
\begin{eqnarray}
\label{eqn:power}
P_{\Phi}(k) &=& P_{\phi}(k) + \mathcal{O}(f^{2}_{nl})\, , \\
\label{eqn:bi}
B_{\Phi}(k_1,k_2,k_{12})&=& 2f_{nl}[P_\phi(k_1)P_\phi(k_2)+{\rm cyclic}]+\mathcal{O}(f_{nl}^3)\, ,
\end{eqnarray}
In practice, we compute the skewness parameter, $S_{3}$, by employing the following approximate formula given in 
\citet{AC12}:
\begin{equation}
\label{eqn:S3_local}
S_{3} = f_{nl}\frac{1.56}{\sigma^{0.84}}10^{-4}\, .
\end{equation}

With the help of the same perturbative technique that \citet{lam-etal09} employed to derive 
Equation (\ref{eqn:plambda_png}), one can straightforwardly show that the conditional probability density in the 
presence of primordial non-Gaussianity of the local type, 
$p_{\rm ng}({\vec \lambda}\vert{\vec \lambda}^{\prime}={\vec \lambda}_{c})$, can be approximated at first order 
as 
\begin{equation}
\label{eqn:pmm_png}
p_{\rm ng}({\vec \lambda}\vert {\vec \lambda}^{\prime}={\vec \lambda}_{c}) = 
\left[1+\frac{\sigma S_{3}}{6}H_3\bigg(\frac{\delta}{\sigma}\bigg)\right]
p({\vec\lambda}\vert{\vec \lambda}^{\prime}={\vec \lambda}_{c})\, .
\end{equation}
Replacing $p({\vec \lambda})$ in Equation (\ref{eqn:plambda}) by $p_{\rm ng}({\vec \lambda})$ in 
Equation (\ref{eqn:plambda_png}) and $p({\vec \lambda}\vert{\vec \lambda}^{\prime}={\vec \lambda}_{c})$ in 
Equation (\ref{eqn:pmm}) by $p_{\rm ng}({\vec \lambda}\vert {\vec \lambda}^{\prime}={\vec \lambda}_{c})$ in 
Equation (\ref{eqn:pmm_png}), one can compute the EZL mass function of dark halos in the presence of 
primordial non-Gaussianity of the local-type.

It is worth explaining here why we restrict our analysis to the local $f_{nl}$ case. When \citet{lam-etal09} 
derived Equation (\ref{eqn:plambda_png}), it was assumed that the only non-vanishing skewness 
is the linear density contrast $\delta\equiv \sum_{i=1}^{3}\lambda_{i}$. This key assumption is valid only for 
the case of primordial non-Gaussianity of the local type which does not modify the shape of the probability 
density distribution of the shear eigenvalues (private communication with T.Y. Lam 2014).
For the other types which depend explicitly on the shape it is expected that the joint probability density of the shear 
eigenvalues will be different from Equation (\ref{eqn:plambda_png}), the derivation of which is beyond the 
scope of this paper.

\subsection{Comparison with N-body Results}\label{sec:test}

To numerically test the EZL mass functions for the local $f_{nl}$ case, we make a use of the samples of dark matter 
halos from a large N-body simulation provided by C. Wagner through private communication. While the detailed 
full description of the N-body simulation can be found in \citet{wagner-etal10} and \citet{WV12}, let us provide 
relevant key information on the numerical data used for the current work: 
\citet{wagner-etal10} utilized the publicly available gadget-2 code \citep{gadget2} to run a $N$-body simulation of 
$1024^{3}$ dark matter particles in a periodic box of linear size $L_{\rm box}=1875\,h^{-1}$Mpc
for a flat $\Lambda$CDM model with non-Gaussian initial conditions with the key cosmological parameters 
set at $\Omega_{m}=0.27,\ \Omega_{b}=0.047,\ h=0.7,\ n_{s}=0.95,\ \sigma_{8}=0.7913$. 
They have identified the dark halos with the Amiga's Halo Finder which calculates the halo mass by considering all particles 
inside a sphere within which a mean overdensity reaches some "redshift dependent" virial overdensity \citep{KK09}.  
From their simulations were produced the samples of the dark halos with mass $M\ge 10^{13}\,h^{-1}M_{\odot}$ at three different 
redshifts ($z=0,\ 0.67,\ 1$) for two different cases of primordial non-Gaussianity of the local type ($f_{nl} = 60,\ 250$) as well as 
for the Gaussian case ($f_{nl}=0$). 

As done in \citet{wagner-etal10}, we determine the numerical mass function of dark halos at each redshift as the differential number 
density of the dark halos per logarithmic mass bins divided by the total volume of the simulation. To estimate the errors associated with 
the determination of the numerical mass function of the dark halos, the sample of the dark halos is divided into eight subsamples, each 
of which has the same size. For each subsample, the number counts are recalculated and then the Jackknife errors per each 
logarithmic mass bin is calculated as the standard deviation scatter among the eight subsamples. 

When \citet{LL13} fitted the EZL mass function to the numerical results to determine its best-fit parameters , the numerical results 
they used were obtained from dark halos identified by the conventional halo-finding algorithms such as the friends-of-friends (FoF) and 
the spherical over density (SO) algorithm. Whereas the dark halos from the $N$-body simulations of \citep{wagner-etal10} used in the 
current work were identified by the Amigo's Halo Finder that is different from the conventional halo finding scheme. Therefore, 
one might expect that the best-fit values of the model parameters of the EZL mass function could be different from those found in the 
original work of \citep{LL13}. Fitting the EZL model to the numerical mass function at $z=0$ for the Gaussian case by adjusting the 
three model parameters,  we determine the best-fit values as $\lambda_{1c}=0.56,\ \lambda_{2c}=0.555,\ \lambda_{3c}=0.32$, which 
turn out to be identical to the best-fit values found by \citep{LL13} for  the case of  the SO halos, indicating the robustness of the EZL 
formula. 

\begin{figure}
\includegraphics[scale=0.8]{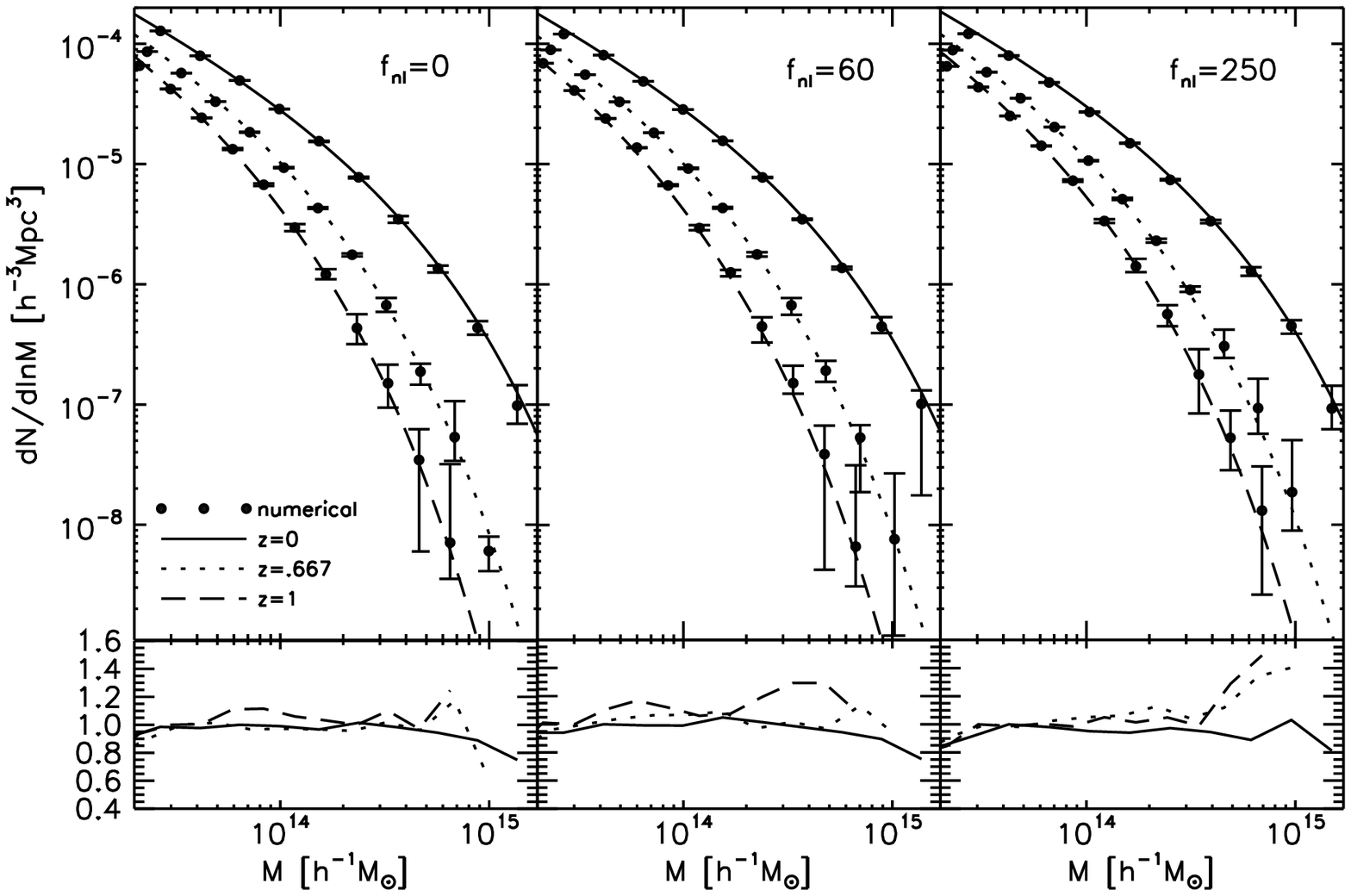}
\caption{(Top panels): Number densities of the cluster-size halos  as dots with Jackknife erros at three different redshifts 
for three different cases of primordial non-Gaussianity of the local type: $f_{nl}=0,\ 60,\ 250$ in the left, middle and right 
panel, respectively. In each panel, the EZL mass function with the best-fit values of  
$\lambda_{1c}=$, $\lambda_{2c}=$, $\lambda_{3c}$ 
are also plotted as solid ($z=0$), dotted ($z=0.66$) and dashed ($z=1$) line, respectively.
(Bottom panel):  Ratios of the EZL mass functions to the corresponding numerical redshifts.}
\label{fig:halo_local}
\end{figure}

The top panels of Figure \ref{fig:halo_local} show the numerical mass functions (dots) with the Jackknife errors as well as the EZL 
mass functions at three different redshifts ($z=0, 0.67$ and $1$ as the solid, dotted and dashed lines, respectively) for three different 
cases of the local non-Gaussianity ($f_{nl}=0,\ 60$ and $250$ in the left, middle and right panel, respectively).  
The same values of the cosmological parameters that were used for the N-body simulations are implemented into the EZL formula, 
and the same values of the EZL model parameters as determined for the Gaussian case, $\lambda_{1}=0.56$, $\lambda_{2}=0.555$
 and $\lambda_{3}=0.32$, are also implemented into the formula for three different cases of $f_{nl}$. 
As can be seen, for all of the three cases of $f_{nl}$, the constant values of the model parameters of the EZL formula 
yield excellent agreements with the numerical results at all redshifts, indicating that the EZL formula with the same parameters 
used for the Gaussian case still works very well even in the presence of primordial non-Gaussianity.

To quantify how good the agreements are, we also show the ratios of the EZL models to the numerical results in the bottom 
panel of Figure \ref{fig:halo_local}. In the mass-range of $M\le 3\times 10^{14}\,h^{-1}M_{\odot}$, the EZL model agrees with 
the numerical mass functions within $20\%$ errors for all cases. In the higher mass section, however, the errors exceed $20\%$.  
The EZL formula exhibit better agreements with the numerical results at lower redshifts and for the case of small $f_{nl}$. 
The rather large deviation of the EZL formula from the numerical result for the case of $f_{nl}=250$ should be due to the fact that 
the EZL mass function with the local-type non-Gaussianity was derived only at first order. 
   
\section{THE EZL MASS FUNCTION OF THE ISOLATED CLUSTERS}\label{sec:iso_ezl}

It is  \citet{SL09} who have done the first feasibility study of using the abundance of the isolated clusters as a probe of 
primordial non-Gaussianity.  Original as the idea of \citet{SL09} was, their analytic prescription of evaluating the 
mass function of the isolated clusters was such a crude approximation based on several oversimplified assumptions. 
In fact, their feasibility study was aimed only at presenting a proof of the concept that the mass function of the isolated 
clusters should be a more sensitive test of the presence of primordial non-Gaussianity than that of all clusters. 

\citet{lee12} constructed a much more accurate formula for the mass function of the isolated clusters for 
the Gaussian case in the framework of the "drifting barrier" (DB) formalism developed by \citet{CA11a,CA11b}. 
\citet{AC12} incorporated the effect of primordial non-Gaussianity into the DB model and confirmed that the 
abundance of the isolated clusters indeed varies more sensitively with the degree of primordial 
non-Gaussianity than that of all clusters. 
True as it is that the DB model of \citet{AC12} is capable of predicting quite accurately the 
abundance of the isolated clusters in the presence of primordial non-Gaussianity,  the model parameters of the DB 
mass function were shown to be not constant against the changes of $z$ and $f_{nl}$ 
\citep{achitouv-etal13b}. 

Noting that the model parameters of the EZL mass function are found to have desirable independence on $z$ 
and $f_{nl}$ in section \ref{sec:test} and given that the dynamical process of the isolated clusters is expected to be 
much simpler than that of ordinary clusters located in over dense regions \citep{des08}, we now attempt to model the 
abundance of the isolated clusters for the local $f_{nl}$ case by employing the following one dimensional (1D) EZL 
formula that has only one parameter other than the overall normalization factor \citep{LL13}:
\begin{equation}
\label{eqn:1dezl}
\int_{\lambda_{3c}}^{\infty}\ d{\lambda}_{3}\,p[\lambda_{3};\sigma(M,z)] \propto 
\int_{\ln M}^{\infty}d\ln M^{\prime}\frac{M^{\prime}}{{\bar\rho}}\frac{dN_{\rm I}(M,z)}{d\ln M^{\prime}}
P(\lambda_{3}\ge {\lambda}_{3c}|\lambda^{\prime}_{3}={\lambda}_{3c})\, ,
\end{equation}
where $dN_{\rm I}/d\ln M$ denotes the mass function of the isolated clusters which have no neighbor 
clusters within a given threshold distance and $A$ is the normalization factor whose value has to be determined 
empirically according to the constraint that the integration of the mass function of the isolated clusters must yield 
the total number of the isolated clusters in a given sample divided by the total volume \citep{lee12}.

The left-hand side of Equation (\ref{eqn:1dezl}) represents the cumulative probability that the smallest shear 
eigenvalue , $\lambda_{3}$, is larger than the characteristic parameter,  $\lambda_{3c}$, 
on the mass scale of $M$. The one-point probability density distribution of $\lambda_{3}$ can be obtained by 
integrating the three-point probability density of ${\vec \lambda}$ as \citep{LS98}
\begin{equation}
\label{eqn:plambda_3}
p(\lambda_{3}) = \int_{\lambda_{3}}^{\infty}d\lambda_{1}\int_{\lambda_{3}}^{\lambda_{1}}d\lambda_{2}
p(\lambda_{1},\lambda_{2},\lambda_{3})\, .
\end{equation} 
The conditional probability, $P(\lambda_{3}\ge \lambda_{3c}|\lambda^{\prime}_{3}=
\lambda_{3c})$, in the right-hand side of Equation (\ref{eqn:1dezl}) can be calculated as 
\begin{equation}
\label{eqn:1dpmm}
P(\lambda_{3}\ge \lambda_{3c}|\lambda^{\prime}_{3}=
\lambda_{3c}) = \frac{\int_{\lambda_{3c}}^{\infty}d\lambda_{3}\
p(\lambda_{3},\lambda^{\prime}_{3}=\lambda_{3c})}
{p(\lambda^{\prime}_{3}=\lambda_{3c})}\, ,
\end{equation}
where $p(\lambda_{3},\lambda^{\prime}_{3})$ denotes the joint probability density distribution of the smallest 
eigenvalues on two different scales, $M^{\prime}$ and $M$, respectively, which can be obtained by integrating 
the six-point probability density of ${\vec\lambda}$ and ${\vec\lambda}^{\prime}$ as \citep{des08}:
\begin{equation}
\label{eqn:jplambda_3}
p(\lambda_{3},\lambda^{\prime}_{3})=\int_{\lambda_{3}}^{\infty}d\lambda_{1}\int_{\lambda_{3}}^{\lambda_{1}}
d\lambda_{2}\int_{\lambda^{\prime}_{3}}^{\infty} d\lambda^{\prime}_{1} 
\int_{\lambda^{\prime}_{3}}^{\lambda^{\prime}_{1}}d\lambda^{\prime}_{2}\ p({\vec \lambda},{\vec \lambda}^{\prime})\, .
\end{equation}
To evaluate the mass function of the isolated clusters in the presence of primordial non-Gaussianity of the local type, 
we replace $p({\vec \lambda})$ in Equation (\ref{eqn:plambda_3}) by $p_{\rm ng}({\vec \lambda})$ and 
$p({\vec \lambda},{\vec \lambda}^{\prime})$ in Equation (\ref{eqn:jplambda_3}) by 
$p_{\rm ng}({\vec \lambda},{\vec \lambda}^{\prime})$, respectively. 

For the case of all clusters, it was shown by \citet{LL13} that the 1D EZL formula characterized by only one 
parameter does not provide a good fit to the numerical results.  However, for the case of the isolated clusters located 
in the under dense regions where the formation process is less affected by the environments, we speculate that the 1D 
EZL formula may work very well as the simpler formation process would be modeled by fewer parameters.  
To test this speculation against N-body simulations, we first construct a subsample of the isolated clusters from 
the N-body data described in section \ref{sec:test}.  \citet{lee12} identified the isolated clusters as those which have 
no neighbor clusters within the distance of $0.4\bar{d}_{c}$ where $\bar{d}_{c}$ denotes the mean separation 
distance of the clusters in the  sample. Basically, we apply the FoF algorithm with the linking length 
parameter of $b=0.4$ to the cluster-size halos with mass larger than $10^{13}\,h^{-1}M_{\odot}$ in the cluster 
sample to find the FoF groups consisting of the clusters. Then, we select the isolated clusters as those FoF groups 
which have only one member cluster. 

The total number  and mean mass of the isolated clusters at $z=0,\ 0.67$ and $1$ are listed in Tables 
\ref{tab:iso_z0}, \ref{tab:iso_z0.6} and \ref{tab:iso_z1}, respectively. As one can see, there are more isolated clusters in the 
models with higher degree of primordial non-Gaussianity. 
Carrying out the same procedure as described in section \ref{sec:test}, we count the number of the isolated clusters 
per logarithmic mass bin at each redshift for each case of the local $f_{nl}$. Then, we fit the 1D EZL formula at $z=0$ 
for the case of $f_{nl}=0$ to the numerical mass functions of the isolated clusters by adjusting the value of 
$\lambda_{3c}$ in the 1D EZL formula, and find that $\lambda_{3c}=0.5$ gives the best-fits. 
Then, plugging this same value of $\lambda_{3c}=0.5$ into Equations (\ref{eqn:1dpmm}), we examine whether 
the same value of $\lambda_{3c}$ makes the 1D EZL formula for the case of non-zero value of $f_{nl}$ match 
the numerical mass functions. 

\begin{figure}
\includegraphics[scale=0.8]{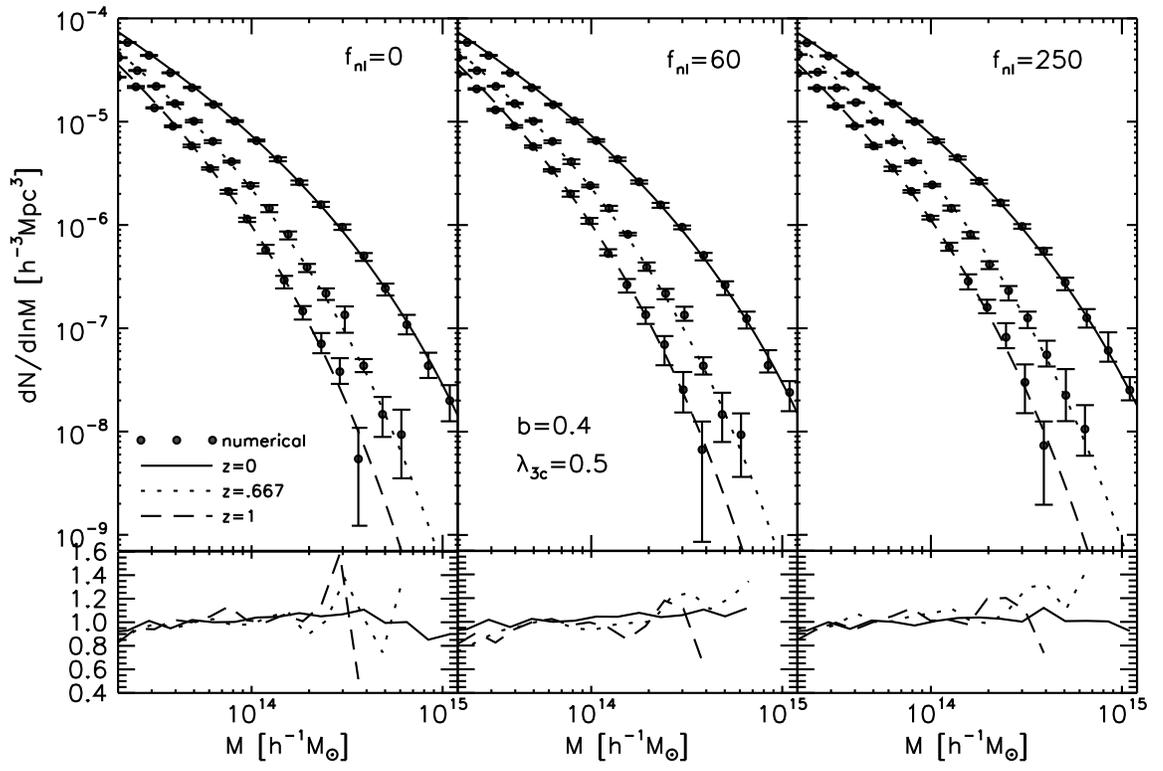}
\caption{Same as Figure \ref{fig:halo_local} but for the case of the isolated clusters.}
\label{fig:iso_local}
\end{figure}

Figure \ref{fig:iso_local} shows the same as Figure \ref{fig:halo_local} but for the cases of the isolated clusters. 
For this plot, we renormalize the 1D EZL mass functions according to the  condition of 
$\int dN_{\rm I}/d\ln M = N_{\rm I,tot}/V$, where $N_{\rm I,tot}$ represents the total number of the isolated clusters found 
in the sample at each redshift for each case of $f_{nl}$ and $V$ is the total volume of the simulation. 
As can be seen, the 1D EZL mass function with the constant value of its model parameter agrees with the numerical results within 
$20\%$ errors in the mass range of $3\times 10^{13}\le M/(h^{-1}M_{\odot})\le  2\times 10^{14}$ for every case. 

\begin{figure}
\includegraphics[scale=0.85]{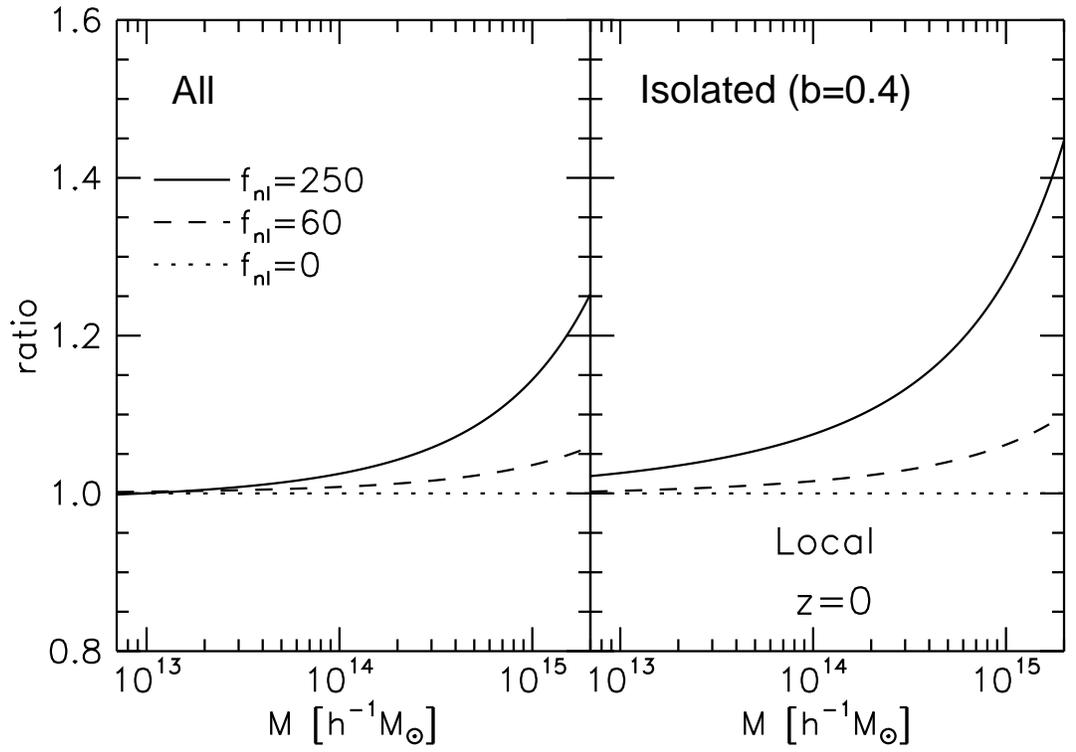}
\caption{Ratios of the mass functions with primordial non-Gaussianity of the local type to the Gaussian ones at 
$z=0$. The left and the right panels corresponds to the cases of all clusters and isolated clusters, respectively.}
\label{fig:localratio_z0}
\end{figure}

To see how sensitively the mass function of the isolated clusters changes with the value of $f_{nl}$, we compute the 
ratio of $[dN_{\rm I}/d\ln M]_{f_{nl}\ne 0}$ to $[dN_{\rm I}/d\ln M]_{f_{nl}=0}$  where 
$[dN_{\rm I}/d\ln M]_{f_{nl}=0}$ and $[dN_{\rm I}/d\ln M]_{f_{nl}\ne 0}$ represent the mass functions of the isolated 
clusters for the Gaussian and the non-Gaussian case, respectively.  
Figure \ref{fig:localratio_z0} plots this ratio in the right panel at $z=0$, while the left panel shows the ratio of 
$[dN/d\ln M]_{f_{nl}\ne 0}$ to $[dN/d\ln M]_{f_{nl}=0}$  where $dN/d\ln M$ is the mass function of all clusters.  

\begin{figure}
\includegraphics[scale=0.85]{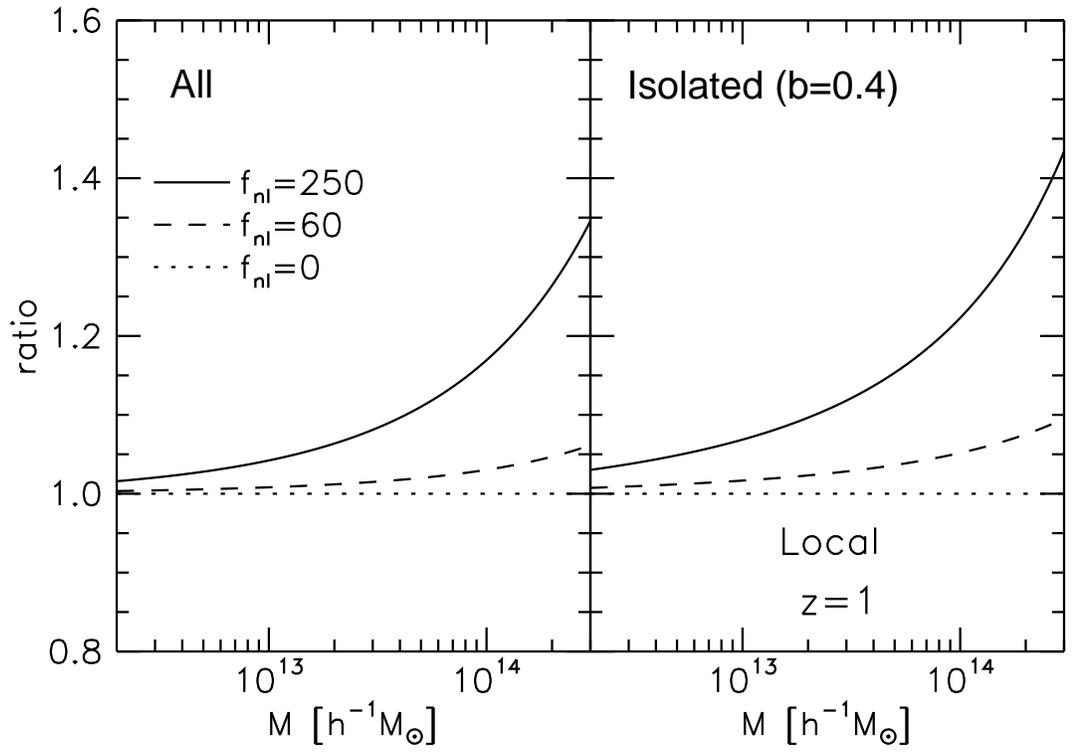}
\caption{Same as Figure \ref{fig:localratio_z0} but at $z=1$.}
\label{fig:localratio_z1}
\end{figure}

Note that the degree of the deviation of the ratio from unity is higher for the case of the isolated clusters than 
for the case of all clusters over the whole mass range.  At the high mass end of $M\ge 2\cdot 10^{15},h^{-1}M_{\odot}$, 
the number counts of the isolated clusters for the case of $f_{nl}=250$ ($f_{nl}=60$) exhibits $\sim 40\%$ ($\sim 10\%$) 
difference from those for the Gaussian case, while in the number counts of all clusters there is only $\sim 20\%$ ($\sim 5\%$) 
difference between the two cases. Figure \ref{fig:localratio_z1} plots the same as Figure \ref{fig:localratio_z0} but at $z=1$, which 
reveals the same trend that the mass function of the isolated clusters in the high-mass end is twice more sensitive to the change 
of $f_{nl}$.  

In practice the EZL mass function of the isolated clusters in the high-mass end would suffer inevitably from much larger 
Poisson errors than that of all clusters  because the formation of massive clusters are strongly suppressed in the isolated 
under dense regions. In our analysis, it is found that the Poisson errors in the measurement of the number densities of the 
isolated clusters in the high-mass range of $10^{14}\le M/(h^{-1}M_{\odot})\le 10^{15}$ are $(2-4)$ times larger than that of 
all clusters. But, Figure \ref{fig:localratio_z0} also reveals that in the low-mass section $10^{13}\le M/(h^{-1}M_{\odot})\le 10^{14}$ 
where the contamination by the Poisson errors is expected to be negligible, 
the ratio of $[dN_{\rm I}/d\ln M]_{f_{nl}=250}$ to $[dN_{\rm I}/d\ln M]_{f_{nl}=0}$, deviates appreciably from unity, much 
higher than the ratio of $[dN/d\ln M]_{f_{nl}=250}$ to $[dN/d\ln M]_{f_{nl}=0}$. Thus, the EZL mass function of the low-mass 
isolated clusters may be useful only for putting an upper limit on the value $f_{nl}$. 

\section{CONCLUSION AND DISCUSSION}\label{sec:con}

We have incorporated primordial non-Gaussianity of the local type into the EZL formula for the halo mass function 
which was originally developed by \citet{LL13} for the case of Gaussian initial conditions. Testing the EZL formula for 
two different cases of the primordial non-Gaussianity parameter ($f_{nl}=60,\ 250$) as well as for the Gaussian case 
($f_{nl}=0$) at three different redshifts ($z=0,\ 0.67,\ 1$) against the numerical results from high-resolution N-body simulations, 
we have found that the EZL mass functions agree excellently with the numerical results for every case considered and 
that its three model parameters have constant best-values, being independent of $z$ and $f_{nl}$. 

We have also constructed the EZL mass function of the isolated clusters which turns out to have only one model parameter 
other than the overall normalization factor. The constant value of the single parameter has been found to yield remarkable 
agreements with the N-body results for all of the three cases of $f_{nl}$ at all of the three redshifts.  Then, we have shown that 
although the abundance of the isolated clusters evaluated by the EZL formula is much more sensitive to the change of the value of 
$f_{nl}$ than that of all clusters in the high-mass end, the practical usefulness of the EZL mass function of the isolated clusters 
is limited to putting an upper limit on the value of $f_{nl}$ due to the relatively large Poisson errors.
 
In the current work, we have restricted our investigation to the case of primordial non-Gaussianity of the local type  
because the joint probability density of the linear shear eigenvalues, 
Equations (\ref{eqn:plambda_png})-(\ref{eqn:pmm_png}),  which are the key quantities in the EZL framework, are valid 
only for the case of local primordial non-Gaussianity. 
Furthermore, our formula is also limited to the case that $f_{nl}$ is not large since Equation (\ref{eqn:plambda_png}) 
is the first order approximation. To use the EZL mass function of the clusters as a probe of primordial 
non-Gaussianity, however, it will be desirable to derive the higher order approximation of the probability density 
distribution of the shear eigenvalues and to find its expression also for the other types of primordial 
non-Gaussianity. 

A more fundamental issue about the EZL mass function is to find a physical meaning of its characteristic model 
parameters. As stated explicitly in \citet{LL13}, the EZL mass function is a mere phenomenological fitting formula 
and thus its characteristic parameters have nothing to do with a collapse condition. In other words, the best-fit 
values of the EZL parameters contain no information on the underlying dynamics that governs real process of the 
gravitational collapse of the density inhomogeneities . A remaining crucial question is why and how the EZL parameters 
stay constant against the changes of redshifts, the key cosmological parameters, and even the primordial non-
Gaussianity parameter in spite of the fact that they are just fitting parameters. 
We plan to work on the above two issues and report the result elsewhere in the future.

\acknowledgments

We are very grateful to C.Wagner for kindly making his simulation data available to us.  We also thank T.Y.Lam 
for helpful discussion. JL was supported by Basic Science Research Program through the National Research 
Foundation of Korea(NRF) funded by the Ministry of Education (NO. 2013004372) and partially by the research grant 
from the National Research Foundation of Korea to the Center for Galaxy Evolution Research  (NO. 2010-0027910).

\clearpage
\clearpage
\clearpage
\clearpage
\clearpage
\begin{deluxetable}{ccc}
\tablewidth{0pt}
\setlength{\tabcolsep}{5mm}
\tablecaption{primordial non-Gaussianity parameter, total number and mean mass of the isolated clusters at $z=0$.}
\tablehead{$f_{nl}$ & $N_{\rm I,tot}$ &$\bar{M}$ \\
& & $[10^{13}\,h^{-1}M_{\odot}]$} 
\startdata
0 & $823903$ & $2.75$\\
60 & $824430$ & $2.76$ \\
250 & $820363$ & $2.78$ \\
\enddata
\label{tab:iso_z0}
\end{deluxetable}
\clearpage
\begin{deluxetable}{ccc}
\tablewidth{0pt}
\setlength{\tabcolsep}{5mm}
\tablecaption{Same as Table \ref{tab:iso_z0} but at $z=0.67$.}
\tablehead{$f_{nl}$ & $N_{\rm I,tot}$ &$\bar{M}$ \\
& & $[10^{13}\,h^{-1}M_{\odot}]$} 
\startdata
0 & $508980$ & $2.22$\\
60 & $512175$ & $2.22$ \\
250 & $516260$ & $2.25$ \\
\enddata
\label{tab:iso_z0.6}
\end{deluxetable}
\clearpage
\begin{deluxetable}{ccc}
\tablewidth{0pt}
\setlength{\tabcolsep}{5mm}
\tablecaption{Same as Table \ref{tab:iso_z0} but at $z=1$.}
\tablehead{$f_{nl}$ & $N_{\rm I,tot}$ &$\bar{M}$ \\
& & $[10^{13}\,h^{-1}M_{\odot}]$} 
\startdata
0 & $352303$ & $1.98$\\
60 & $356343$ & $1.99$ \\
250 & $364036$ & $2.02$ \\
\enddata
\label{tab:iso_z1}
\end{deluxetable}

\end{document}